\documentstyle[amssymb,aps,twocolumn]{revtex}

\newcommand{\beq}{\begin{equation}}
\newcommand{\eeq}{\end{equation}}
\newcommand{\beqa}{\begin{eqnarray}}
\newcommand{\eeqa}{\end{eqnarray}}

\begin{document}
\title{Faint laser quantum key distribution: Eavesdropping exploiting multiphoton
pulses}
\author{St\'{e}phane F\'{e}lix, Nicolas Gisin, Andr\'{e} Stefanov and Hugo Zbinden*.}
\address{{\small {\em Group of Applied Physics, University of Geneva, 1211 Geneva 4, }%
}Switzerland\\
hugo.zbinden@physics.unige.ch, phone: +41 22 702 68 83, fax: +41 22 781 09\\
80}
\date{\today}
\maketitle

\begin{abstract}
The technolgical possibilities of a realistic eavesdropper are discussed.
Two eavesdropping strategies taking profit of multiphoton pulses in faint
laser QKD are presented. We conclude that, as long as storage of Qubits is
technically impossible, faint laser QKD is not limited by this security
issue, but mostly by the detector noise.
\end{abstract}

\section{Introduction}

Quantum Cryptography or Quantum Key Distribution \cite{BB84,Gisin2001}
offers provably secure key exchange\cite{Hitoshi}. Even for unlimited
technological power of the eavesdropper, Eve, and technically limited
devices of the legitimated users, Alice and Bob, the privacy of the
exchanged key can be guaranteed.

In principle, single photons are sent from Alice to Bob. A quantum degree of
freedom (e.g. polarization) is used to encode a bit value (0 or 1). In
practice, single photons are often mimicked by faint laser pulses containing
on average much less than one photon per pulse. Hence, multiple photon
pulses occur and open the door to efficient eavesdropping strategies that
threaten the security of Quantum Key distribution, especially in combination
with high losses in the fiber link \cite{Lütkenhaus}. In particular, suppose
that:

i) due to losses in the fiber link, the fraction of pulses arriving at Bob
links is smaller than the fraction of pulses containing more than one photon
leaving Alice,

ii) Eve possesses a lossless fiber link from Alice to Bob,

iii) Eve can measure the number of photons without perturbing them
(QND-measurement) and

iv) Eve can store photons without perturbing them for minutes.\newline
In this case, Eve could measure the number of photons of the pulse. If there
are less than two, she simply blocks the pulse. If there are more than one
photon, she stores one of them and sends the other ones to Bob. After Bob
and Alice announced the bases of the measurements, Eve measures her photons
in the right bases and gets the full information. She does not create any
errors and the average number of photons arriving at Bob remains unchanged.

In this paper we would like to discuss the security of faint laser QKD. We
put emphasize on the multi-photon problem, considering realistic set-ups
with nowadays technical limits and realistic eavesdroppers with state-of-art
or near-future technology, but no unlimited technological power. For
simplicity, we will restrict ourselves to the well-known 4-states protocol
BB84\cite{BB84}.

As you can imagine, the points ii) to iv) above are not easy to realize.
There are no lossless fibers, no efficient QND-measurements and no efficient
storage of photons for more than $\mu $s\ nor transfer of the quantum degree
of freedom to another quantum system like an ion in a trap. In section III
we define the techniques a realistic Eve might possess. Nevertheless, using
almost feasible techniques efficient eavesdropping is possible. Acting on
two-photon pulses, Eve has two possibilities: First, she can apply an
intercept-resend strategy making a measurement on both photons and sending a
new photon to Bob according to the obtained result. In section IV we develop
this strategy and calculate Eve's information per created bit error. Second,
she can measure one photon and let pass the other one unchanged to Bob. In
this case she won't introduce any errors, but she alters the photon
statistics at Bob. In section V we calculate Eve's information as a function
of the number of 2-photon pulses at Bob. In section VI, finally, we discuss
the consequences of these eavesdropping strategies on the privacy
amplification, the net bit rate and the maximal transmission distances.

In the next section we sum up the basics equations giving bit rates and
photon number probabilities.

\section{Transmission rates and photon statistics.}

Alice produces coherent states with a low mean photon number $\mu $ which
are only an approximation of single photon Fock-states. The probability that
one finds $n$ photons in a coherent state follows the Poisson statistics: 
\begin{equation}
P(n,\mu )=\frac{\mu ^{n}}{n!}e^{-^{\mu }}
\end{equation}
In a second order approximation we obtain:

\[
P(0)\approx 1-\mu +\frac{\mu ^{2}}{2} 
\]
\begin{equation}
P(1)\approx \mu -\mu ^{2}
\end{equation}
\begin{equation}
P(2)\approx \frac{\mu ^{2}}{2}
\end{equation}
Accordingly, the probability that a non-empty pulse contains more than 1
photon becomes: 
\begin{equation}
P(n\geq 2|n>0)=\frac{1-P(0)-P(1)}{1-P(0)}\approx \frac{\mu }{2}+\left( \frac{%
\mu ^{2}}{4}\right)
\end{equation}
The probability to have at least one photon is $1-P(1)=1-e^{-\mu }$.
Consequently, the probability to have a (single) count at Bob is: 
\begin{equation}
P_{single}=1-e^{-^{\mu t_{ab}\eta _{b}}}\approx \mu t_{ab}\eta _{b}
\label{Rbob}
\end{equation}
where $\mu $ has been replaced by $\mu t_{ab}\eta _{b}$, with $t_{ab}$ the
transmission of the optical fiber and $\eta _{b}$ the detection efficiency
of Bob's photon counter. The raw bit rate (before sifting, error correction
and privacy amplification) is simply 
\begin{equation}
R_{raw}=\upsilon P_{single}\approx \upsilon \mu t_{ab}\eta _{b}
\end{equation}
with the pulse frequency $\nu $.

We can express t$_{ab}$ in terms of the total loss $L_{ab}$ (in dB), the
product of $\alpha _{ab}$ (in dB/km) and $l_{ab}$, the length of the fibre
in km.

\begin{equation}
t_{ab}=10^{-\frac{L_{ab}}{10}}=10^{-\frac{\alpha _{ab}l_{ab}}{10}}
\end{equation}
Eve's attacks must leave $P_{single}$ unchanged. Eve can make a
beam-splitting attack (coupling out a fraction $\lambda $) and increase the
fiber transmission up to $t_{e}$. Thus, she obtains:

\begin{equation}
\lambda t_{e}=\lambda 10^{\frac{-\alpha _{e}l_{e}}{10}}=10^{\frac{-\alpha
_{ab}l_{ab}}{10}}=t_{ab}
\end{equation}

Bob has two detectors per basis. Whenever he measures in a non-compatible
basis, he has a certain probability to obtain a coincidence count. In the
non compatible basis, the polarizing beamsplitter acts like a normal 50\%
coupler and we have on average $\frac{\mu }{2}t_{ab}$ photons in each arm.
Accordingly the probability to have a click in both detectors becomes (the
factor $\frac{1}{2}$ is for the probability to choose the wrong basis): 
\begin{equation}
P_{coinc}=\frac{1}{2}\left[ 1-e^{-^{\frac{\mu }{2}t_{ab}\eta _{b}}}\right]
^{2}\approx \frac{1}{8}\mu ^{2}t_{ab}^{2}\eta _{b}^{2}\approx \frac{1}{4}%
P(2)t_{ab}^{2}\eta _{b}^{2}  \label{BobCoi}
\end{equation}
This is in first approximation equal to the intuitive calculation, taking
simply the product of the probabilities to have two photons $P(2)$, the
probability they arrive both at Bob $t_{ab}^{2},$ the probability that the
non-compatible basis is chosen $\frac{1}{2}$, the probability that photons
do not go to the same detector $\frac{1}{2}$, and finally the probability
that both are detected $\eta _{b}^{2}$. (In the case of passive choice
between the two bases, we find a factor of $\frac{5}{8}$ instead of $\frac{1%
}{4}).$ Again, Eve must make sure that due to her intervention the number of
coincidences is not changed.

\section{The technology of a ''realistic'' Eve}

In this paragraph, we present the technology at the disposal of a
''realistic'' Eve. We consider technolgies that are either state of the art
or are imaginable with unlimited financial possibilities in the near future.
It is important to note, that Eve can only take profit from the technologies
available at the moment of key exchange. This is in contrast to the public
key system, where messages that should remain secret for a long time are
also threatened by future technologies. For simplicity, in some cases when
there is no important consequence for the efficiency of her attacks, we
attribute to Eve perfect technology.

\subsection{Eve's infrastructure and optical components}

Eve has free access to Alice and Bob's quantum channel at any point outside
their offices. She can install optical components (switches, couplers etc.)
without being noticed. Eve's components have no losses and no backscattering
and are therefore invisible. Her optical switches are arbitrarily fast. In
particular, she can in arbitrary short times recover the original situation
whenever Alice and Bob check their line with an OTDR (optical time domain
reflectometer). She knows Alice and Bob's bases and can perfectly align her
analyzers with respect to these bases\footnote{%
This is trivial in the cases where Alice and Bob use strong classsical
pulses to align their setups. In the auto-aligned ''plug\&play'' setup\cite
{greg2000b}, the phase difference between the two pulses is constantly
fluctuating due to thermal instabilities. Bob could even introduce a random
phase for each bit, rendering the task of Eve even more difficult.}.

\subsection{Eve's photon counters}

Eve's photon counters are perfect. They feature a quantum efficiency of 1
and no noise. Moreover, they can count photons arriving at the same time.
Hence, Eve can in particular distinguish between pulses containing one and
two photons, but the measurement is destructive (no QND).

\subsection{Eve's fibre link}

Eve doesn't have lossless fibres. Nowadays telecom fibres have reached a
such a high quality level that the losses in the second telecom window
(@1550 nm) are close to the theoretical limit. This means that a value of
0.20 dB/km corresponds to the losses you expect from inevitable Rayleigh
backscattering due to the different molecules (essentially silica and
germanium) present in the glass. Fibers with a pure silica core have lower
losses (at the moment down to 0.171 dB/km \cite{Kato}) but are actually not
telecom standard. In contrary, real fibres have additional losses due to
splices and connections in the centrals and a value of 0.25 dB/km is more
realistic. Moreover, installed fibres rarely follow the bee-line ($d_{ab}$)
(they often follow roads and go from one telecom central to another). We
assume that Eve possesses a fiber link relating Alice and Bob on a straight
line with a loss $\alpha _{e}$ of 0.15 dB/km. Alice and Bob must measure the
total loss $L_{ab}$ of their link. Eve's maximum transmission gain $G_{t}$ ($%
G_{t}=10log(\frac{t_{e}}{t_{a}})$) is then $L_{ab}-\alpha _{e}d_{ab}$. If
the time of flight of the photons is monitored, $G_{t}$ is only $(\alpha
_{ab}-\alpha _{e})l_{ab}$, which is according to our assumptions $\approx $%
0.1 dB per fibre km. At any other wavelength $G_{t}$ would be larger.

Note that Eve could introduce in advance additional loss between Alice and
Bob. Crypto users must therefore calculate the loss $L_{ab}$ and $\alpha
_{ab}$ of their line from the obtained $P_{single}$ and compare it to the
usual values$\footnote{%
Note that a standard measurement of the fiber transmission could be
manipulated by Eve.}$. The higher the accepted loss is, the higher $G_{t},$
and as we will see, the higher Eve's information.

\subsection{Eve's photon sources}

Eve possesses perfect single photon sources and more general sources that
can emit arbitrary states at demand. Nowadays, such kind of sources are far
from being realized. However, admitting such sophisticated sources has not
much impact on the security and performance of practical QKD systems, since
in most cases only the single photon and two-photon statistics can be
measured by Bob.

\subsection{Eve's possibilities to modify and influence Alice and Bob's
apparatuses}

In cryptography in general, Alice and Bob's offices have to be supposed to
be secure. We also assume that Eve cannot manipulate Alice and Bob's crypto
devices before they get installed (or that they are quantum physicist that
can test them). This is not trivial, and in practice crypto users have to
trust their suppliers. However, due to the optical fiber coming into the
offices Alice and Bob have to be aware of, what is sometimes called ''Trojan
horse'' attacks. In such kind of attacks Eve sends light pulses at arbitrary
wavelengths and analyzes the reflected light in order to find out e.g. which
detector clicked or which phase was applied. We assume that Alice and Bob
applied all precautionary measures to prevent such strategies. In the same
line, we take it for granted that Alice and Bob do not send any parasite
signals that could reveal their bits. The problem is not that these counter
measures are technically difficult to realize, but the prove that all leaks
have been plugged up.

All these assumptions are essential in the sense that otherwise secure QKD
is not possible. Another assumption commonly made is that all bit errors
detected by Alice and Bob are introduced by Eve, and that she possesses the
corresponding information of the key. This means that privacy amplification
has to be applied to reduce this potential information, which in turn, in
the cases of large quantum bit error rates (QBER), drastically reduces the
bit rate. In practice also without any eavesdropping, QKD-systems suffer
from considerable QBER, that has two origins \cite{hugo}. A first part, QBER$%
_{opt}$, is due to non perfect alignment of the optical setup e.g. the
polarization measurement bases of Alice and Bob are not perfectly parallel,
or the light pulses are not perfectly polarized. The second, and mostly
dominant, part is due to darkcounts of the photon counters. Attributing all
QBER\ to the presence of an eavesdropper and assuming that Eve gains the
corresponding information (see section III), means supposing that Eve has
ways to reduce QBER$_{opt}$ and QBER$_{det}.$ Only diminishing these errors
allows her to tap the line and gain information without introducing
additional errors. On the one hand, one can imagine that Eve can reduce QBER$%
_{opt}$ in certain cases, by improving e.g. the alignment of Alice and Bob's
bases. One the other hand, reducing QBER$_{det}$ i.e. the dark count rate at
a distance by sending some magic light pulses is not conceivable. Therefore,
we suggest that only for the difference between the measured QBER (QBER$%
_{mes}$) and the presumed QBER$_{det}$ privacy amplification must be
applied. Typically, QBER$_{opt}$ is in the order 0.5\%. Thus, adding a
tolerance of 0.5\%, we may attribute an error of 1\% to Eve.

Similarly, we assume that Eve cannot increase the detection efficiencies of
Bob's detectors. This would be very useful for the eavesdropping strategy
discussed in section IV. Instead of lowering the losses in the quantum
channel, Eve could then simply increase the efficiency of the detectors (for
the pulses, from which she was able to extract one photon), as suggested by
L\"{u}tkenhaus\cite{Lütkenhaus}.

\section{Strategy A: Intercept-resend with multi-photon pulses}

Eve intercepts all photons and analyzes them with the setup shown in Fig. 1.
She transmits the measurement results through a (lossless) classical channel
to a source close to Bob that sends the most suitable states, respecting the
photon statistics expected by Bob. Eve sends the photons via passive
beamsplitter to both measurement bases. Consequently, she obtains four
different possible results (cases with more than 2 photons are neglected).
In table 1 the four different possibilities are listed with the
corresponding probabilities p (the conditional probability, provided that at
least one photon is detected), Eve's information per bit $I(A,E)$, the QBER
she creates and the ratio of information per QBER. The case A (one photon)
corresponds to the standard intercept-resend on single photon pulses\cite
{B92}. Case B, two photons, each in a different basis, is the most favorable
one for Eve. First, after the sifting process she will possess the full
information on the bit value. Second, by sending an intermediate state, a
linear combination of the two corresponding states separated by $\frac{\pi }{%
4},$ she will create a smaller error ($0.15$) than in case A. Case C, both
photons in the same detector, is still favorable. In contrary, Eve will
discard if possible all events D, two photons is the same (wrong) basis,
where she gains no information at all.

\[
\begin{tabular}{|c|c|c|c|c|}
\hline
$case$ & $p$ & $I(A,E)$ & $QBER$ & $\frac{I(A,E)}{QBER}$ \\ \hline
\multicolumn{1}{|l|}{A} & $1-\frac{\mu }{2}$ & $\frac{1}{2}$ & $\frac{1}{4}$
& $2$ \\ \hline
\multicolumn{1}{|l|}{B} & $\frac{1}{2}\frac{\mu }{2}$ & $1$ & $sin^{2}\left( 
\frac{\pi }{8}\right) =0.15$ & $6.83$ \\ \hline
\multicolumn{1}{|l|}{C} & $\frac{3}{8}\frac{\mu }{2}$ & $\frac{2}{3}$ & $%
\frac{1}{6}$ & $4$ \\ \hline
\multicolumn{1}{|l|}{D} & $\frac{1}{8}\frac{\mu }{2}$ & $0$ & $\frac{1}{2}$
& $0$ \\ \hline
\end{tabular}
\]
Of course, Eve has to make sure that the number of photons arriving at Bob
remains unchanged. If the transmission of Alice and Bob's quantum channel is
lower than the probability of getting case B, Eve will only send photons in
this case. If this outcome is not abundant enough, she will have to add
pulses after detections of kind C and finally of kind A and possibly even B.
This is illustrated in Fig. 2, where Eve's information per bit error is
given as a function of fiber length. One can notice that from a distance of
about 65 km on, Eve can entirely rely on case B and obtains therefore 3.4
times more information per bit error than in the case A (usual
intersept-resend) that applies in the limit of very short fibers.

Eve pays attention to reproduce the same number of coincidences at Bob,
which is no problem according paragraph III D.

\section{Strategy B: Measuring one photon out of two in a pulse}

The easiest way to take profit of two photons in a pulse is to couple a
fraction $\lambda $ out of each pulse with a beamsplitter and replace the
original line with a line that has lower loss, in order to compensate for $%
\lambda $ (see Fig. 3). Then clearly, the number of photons (and their
statistics) arriving at Bob remains unchanged and this attack does not
generate any errors. Nevertheless, Eve gains the following amout of
information: 
\begin{equation}
I(A,E)=\frac{\mu }{2}\left( 2\lambda \left( 1-\lambda \right) \right) \frac{1%
}{2}  \label{Ieve-B}
\end{equation}
where $\frac{\mu }{2}$ is the probability to find two photons in a pulse, $%
2\lambda 1-\lambda $ is the probability that these two photons separate at
the coupler (outcoupling probability of $\lambda $) and finally the
outcoupled photon has a probability of $\frac{1}{2}$ to end up in the right
basis. We can see immediately that this information is maximum ($I(A,E)=%
\frac{\mu }{8}$) for $\lambda =\frac{1}{2}$, corresponding to a gain $G_{t}$
of 3 db. If (instead of only one beamsplitter) many couplers and analyzers \
are used in series, with a total coupling loss that equals $G_{t}$, the
information can go up to $\frac{\mu }{4}$. This value can be approached for $%
G_{t}\gtrsim 6db$.

In order to further increase her information, Eve must favor the detection
of two photon pulses. She introduces therefore a shutter (see Fig. 3) that
allows her to block a fraction ($1-\gamma )$ of the pulses, when she didn't
catch one photon. In this way the photon distribution is no longer
Poissonian. The calculation of the photon distribution at the entrance port
of Bob, as a function of $\lambda $ and ($1-\gamma ),$ is straightforward
but lengthy. The probabilty for pulses containing n photons (n%
\mbox{$>$}%
0) is

\begin{equation}
P^{\prime }(n,\lambda ,\gamma )=\frac{1}{n!}\left( 1-\lambda \right) ^{n}\mu
^{n}t_{e}^{n}\left[ 
\begin{array}{c}
\left( \gamma -1\right) e^{-\mu +\left( 1-\lambda \right) \left(
1-t_{e}\right) \mu } \\ 
+e^{-\mu \left( 1-\lambda \right) t_{e}}
\end{array}
\right]
\end{equation}
with the transmission of Eve's fiber $t_{e}$. With $P_{single}^{\prime
}(\lambda ,\gamma )\approx \eta _{b}P^{\prime }(1,\lambda ,\gamma )$ we
obtain for Bob's singles count probability : 
\begin{equation}
P_{single}^{\prime }(\lambda ,\gamma )\approx \left( 1-\lambda \right) \mu
t_{e}\eta _{b}\left[ 
\begin{array}{c}
\left( \gamma -1\right) e^{-\mu +\left( 1-\lambda \right) \left(
1-t_{e}\right) \mu } \\ 
+e^{-\mu \left( 1-\lambda \right) t_{e}}
\end{array}
\right]
\end{equation}

This formula corresponds approximatively, to what we obtain by simply adding
up the probabilities for 1- and 2-photon pulses: 
\begin{eqnarray}
P_{single}^{\prime }(\lambda ,\gamma ) &\approx &t_{e}\eta _{b}\left[ 
\begin{array}{c}
2\left( 1-\lambda \right) \lambda P(2) \\ 
+\gamma \left( \left( 1-\lambda \right) P(1)+2\left( 1-\lambda \right)
^{2}P(2)\right)
\end{array}
\right] \\
&\approx &\left( 1-\lambda \right) \mu t_{e}\eta _{b}\left[ \lambda \mu
+\gamma \left( \left( 1-\mu \right) +(1-\lambda )\mu \right) \right]
\end{eqnarray}

The first term (inside the square brackets) is the favorable case, when
there are two photons in the pulse and one of them goes to the detectors.
The second and the third term correspond to 1 and 2 photons propagating to
Bob, respectively. This value must equal to $P_{single}$\ given in eq. [\ref
{Rbob}] without spy. which is the e.g. case for $\gamma =1$ when $t_{ab}=%
\frac{t_{e}}{2}$ and $\lambda =\frac{1}{2}$. This case correspond to the
simple attack with a 3db coupler, presented at the top of this section.

Eve can block all pulses without detection ($\gamma =0),$ as soon as $%
t_{ab}\leq \frac{t_{e}\mu }{4}$. This is the case for $\mu =0.1$ when $%
G_{t}\geq 16dB$. Then, Eve get's 50\% of the information without creating
errors. This simple attack is almost as efficient as an attack based on a
complicated QND measurement, where same result is obtained for $t_{ab}\leq 
\frac{t_{e}\mu }{2}$. If Eve had at disposition some device for storing
photons, she could wait for the sifting process, then measure her photons in
the right basis and get 100\% information. In consequence, this kind of
attacks are very powerful as soon Eve's $G_{t}$ is large.

However, by acting selectively on two-photon pulses Eve changes the photon
statistics and her presence might be revealed by Bob's coincidence
measurements. According to eq.[\ref{BobCoi}] Bob's coincidence rate is $%
P_{coinc}\approx \frac{1}{8}\mu ^{2}t_{ab}^{2}\eta _{b}^{2}$. On the other
hand with $P_{coinc}^{\prime }(\lambda ,\gamma )\approx \frac{1}{4}\eta
_{b}^{2}P^{\prime }(2,\lambda ,\gamma )$ we get: 
\begin{equation}
P_{coinc}^{\prime }=\frac{1}{8}\left( 1-\lambda \right) ^{2}\mu
^{2}t_{e}^{2}\eta _{b}^{2}\left[ 
\begin{array}{c}
\left( \gamma -1\right) e^{-\mu +\left( 1-\lambda \right) \left(
1-t_{e}\right) \mu } \\ 
+e^{-\mu \left( 1-\lambda \right) t_{e}}
\end{array}
\right]
\end{equation}
In order to fulfill the two conditions 
\begin{eqnarray}
P_{single}^{\prime } &=&P_{single} \\
P_{coinc}^{\prime } &=&P_{coinc}  \nonumber
\end{eqnarray}
Eve can vary the parameters $\gamma $ and $\lambda $. But we realize that
for $\gamma \neq 1$, Eve cannot satisfy both conditions. For equal single
counts, she always increases the number of coincidences detected by Bob. If
the effect of Eve's interaction was to reduce the coincidence count rates,
she could add from time to time photons at price of additional errors. But
reducing the number of coincidences without changing the bit rate, is
impossible with linear optics (i.e. without QND-measurement of the photon
number). However, the number of detected coincidences is small and
statistical fluctuation are large. In Fig. 4 we see Bob's coincidence counts
as a function of $\gamma $, for 10$^{10}$ pulses sent over 60 km. The
horizontal lines are the coincidence rates without Eve ($P_{coinc}$), and $%
\pm 2\sigma $ ($P_{coinc}\pm 2\sqrt{P_{coinc}}$) variations. In the same
graph Eve's information is plotted, too. We can readout the information Eve
obtains, by increasing Bob's coincidence counts by 2 $\sigma .$ $I(A,E)$is
given by 
\begin{eqnarray}
I(A,E) &=&\gamma \frac{\mu }{2}\frac{1}{2}2\lambda \left( 1-\lambda \right)
+\left( 1-\gamma \right) \frac{1}{2} \\
&=&\gamma \frac{\mu }{2}\lambda \left( 1-\lambda \right) +\left( 1-\gamma
\right) \frac{1}{2}  \nonumber
\end{eqnarray}
where the two terms are Eve's information for a non-activated shutter
(according eq. (\ref{Ieve-B}) ) and for an activated shutter ($\frac{1}{2}$%
), respectively.

\section{Impact of multiphoton eavesdropping on bitrates and maximum
transmission distance}

After the sifting process Alice and Bob correct the errors in their raw key
and apply privacy amplification in order to reduce Eve's information to an
arbitrarily low level. The efficiency of these processes depends of course
on the QBER and I(A,E), respectively\cite{Brassard2, Bennett}. So high error
rates have their consequence on the net bit rate $R_{net}$ (or distilled bit
rate $R_{dist}$), which is at the end, together with the transmission
distance, the the only figure of merit of QKD. Therefore it is helpful to
plot $R_{net}$ as a function of fiber length. $R_{net}$ decreases in a first
time exponentially (linearly in log plot) due to the fiber losses, before at
a certain point it rapidly falls to zero due to the exploding losses due to
the necessary error correction and privacy amplification. This point,
representing the maximum transmission distance, depends now on the
estimation of Eve's information corresponding to her technical
possibilities. In practice, nowadays systems are essentially limited by the
detector noise, which leads to high QBER at large distances. In order to
make the impact of the discussed eavesdropping strategies more transparent,
and to estimate the limits of future systems with improved detectors, we
assume in the following graphs a dark count probability of 10$^{-6}$ instead
of the actually typical 10$^{-5}$\cite{stucki2001}. We compare the impact of
the presented strategies with the scenario of an Eve with unlimited
technological power as proposed by L\"{u}tkenhaus. The L\"{u}tkenhaus curve
is dropping down rapidly, because whenever $t_{ab}\eta _{b}\approx \frac{\mu 
}{2}$ Eve's information $I(A,E)\approx 1$ and almost everything of the key
is lost by privacy amplification. Therefore the curve is calculated with an
optimal $\mu $ which decreases with the transmission distance (loss). This
has in turn the same effect, that the $R_{net}$ is dropping down. .

In Figure 5, we see $R_{net}$ as a function of distance for strategy A for
different values of $\mu $. We assume Eve's induced error rate to be $%
QBER_{mes}-QBER_{det}\approx 1\%,$ and $I(A,E)$ corresponding to this error
has to be corrected. The maximum transmission distance is not considerably
reduced with respect to an analog curve without multi-photon attacks, or to
the shown curve without eavesdropper (no privacy amplification). The reason
is that in the worst case (case B) we have to correct 6.8\% of information
instead of 2\% for single photon intercept-resend. This means we will
approximately loose 6.8\% of the key instead of 2\%. However, the bit loss
due the error correction is in comparison much more important. So we can say
for strategy A, multiphoton-pulses do not reduce the maximum transmission
loss, nevertheless we have to be aware of the problem and apply the
corresponding privacy amplification.

Figure 6 shows the result for strategy B. We suppose that $I(A,E)$
corresponding to a 2$\sigma $ increase of $P_{coinc}^{\prime }$\ has to be
corrected. The difference with respect to the shown curve without
eavesdropping is small\footnote{%
This would be the same for lossless fibers ($\alpha _{e}=0dB$) since $\gamma 
$ is not limited by the $G_{t\text{ }}$but by the change in $%
P_{coinc}^{\prime }$ (see Fig. 4).}. $R_{net}$ is not considerably reduced
since the maximum $I(A,E)$ is $\frac{1}{2}$, which reduces $R_{net}$ by a
factor\footnote{%
QKD-systems that do not monitor $P_{coinc}$ must systematically correct for
the maximum $I(A,E)$ for a given distance. The fact of measuring the $%
P_{coinc}$\ can increase $R_{net}$ by a factor of up to 2.} of 2. This is
illustrated by the curve where photon storage is admitted and $I(A,E)$ can
go up to 1, and therefore $R_{net}$ drops down faster. However, the
transmission distance is only dramatically reduced, if photon storage is
combined with conservation of photon statistics, which asks for
QND-measurements. Therefore under realistic conditions, even if strategy B
is more efficient than strategy A, it does not limit the transmission of
faint laser QKD.

Our calculation give for both eavesdropping strategies higher $R_{net}$ for
higher $\mu .$ For strategy B, $I(A,E)$ decreases with a increasing $\mu $,
since $\frac{2\sqrt{P_{coinc}}}{P_{coinc}}$ decreases and Eve's presence is
discovered faster. However, our calculation for strategy A is only correct
for $\mu \ll 1$, indeed, it becomes much more powerful for $\mu \approx 1.$
We plan to study the efficiency of this strategy for larger $\mu $, in order
to optimize $\mu $ and hence the bit rate of faint laser QKD.

\section{Conclusions}

We have shown that also with realistic technologies an eavesdropper can
easily take profit of multi-photon pulses. However, the impact on the
maximum transmission distance is minor. In order to guarantee this, Bob must
survey the length of the fiber, its loss and the coincidence count rates in
order to limit Eve's information. We think it is not reasonable to attribute
all QBER to Eve. Along the same line, in the case of increased QBER, it
might be better to check the causes, instead of simply perform error
correction and privacy amplification.

It is important to note, that QND-measurements alone, does not allow
dangerous multi-photon attacks. The crucial point is whether Eve can store
photons or not. One can take for granted that it is impossible to store
millions of qubits efficiently for seconds in the near future. Users who
worry about that may just wait a few minutes before proceeding with sifting.
Therefore for the time being, faint laser QKD has no security problem due to
multiphoton pulses. It is limited in distance by the dark counts of the
detectors. In conclusion, QKD using entangled photons is not mandatory\cite
{greg2000a,Brassard}\footnote{%
Note, that QKD-setups based on parametric downconversion face an identical
multi-photon problem, if the choice of the bases is not passive\cite
{greg2000a}.}.

\section{Acknowledgments}

We thank Norbert L\"{u}tkenhaus and Fran\c{c}ois Cochet for helpful
discussions. This work was supported by the Esprit project 28139 (EQCSPOT)
through Swiss OFES and the FNRS.

\newpage

\section{Figure Captions}

Fig 1: Eve's setup for strategy A: S = optical switch, PBS= polarization
beam splitter, C = photon counter

Fig 2: Strategy A: Eve information $I(A,E)$ according to the different
regimes (see table 1) as function of distance, for $\alpha _{ab}=0.25dB/km$
and $\mu =0.1$.

Fig 3: Eve's setup for strategy B. S = optical switch, PBS= polarization
beam splitter, C = photon counter

Fig. 4: Strategy B: Eve's information $I(A,E)$ and number of coincidences as
a function of $\gamma ,$ under the condition of unchanged single counts ($%
P_{single}^{\prime }=P_{single}$). The parameters are: $\mu =0.1$, number of
pulses 10$^{10},$ $l_{ab}=60km$, $\alpha _{ab}=0.25dB/km$, $\alpha
_{e}=0.15dB/km$

Fig. 5: Strategy A: Normalized $R_{net}$ as a function of distance for
different values of $\mu $. For comparison curves with no eavesdropper (only
error correction applied) and an eavesdropper with unlimited technology
(L\"{u}tkenhaus) are shown.

Fig. 6: Strategy B: Normalized $R_{net}$ as a function of distance. The
parameters are: $\mu =0.1$, number of pulses 10$^{10},$ $l_{ab}=60km$, $%
\alpha _{ab}=0.25dB/km$, $\alpha _{e}=0.15dB/km$. For comparison curves with
no eavesdropper (only error correction applied) an eavesdropper with storage
device and an eavesdropper with unlimited technology (L\"{u}tkenhaus) are
shown.


\section{References}

\begin{references}
\bibitem{BB84}  Ch.H. Bennett, and G. Brassard, ''Quantum cryptography:
public key distribution and coin tossing'', Int. conf. Computers, Systems \&
Signal Processing, Bangalore, India, December 10-12, 175-179 (1984).

\bibitem{Gisin2001}  N. Gisin, G. Ribordy, W. Tittel, and H.Zbinden,
''Quantum Cryptograpy'', quant-ph/0101098, submitted to Rev. Mod. Phys.
(2000).

\bibitem{Hitoshi}  D. Mayers, ''Unconditional security in quantum
cryptography'', Journal for the Association of Computing Machinery (to be
published)(1998); also in quant-ph/9802025; H.-K. Lo and H.F. Chau,
''Unconditional security of quantum key distribution over arbitrary long
distances'' Science {\bf 283}, 2050-2056 (1999); P.W. Shor, and J. Preskill,
''Simple proof of security of the BB84 Quantum key distribution protocol'',
Phys. Rev. Lett. {\bf 85}, 441 (2000)

\bibitem{Lütkenhaus}  N. L\"{u}tkenhaus, ''Security against individual
attacks for realistic quantum key distribution'', Phys. Rev. A, {\bf 61},
052304 (2000).

\bibitem{Kato}  T.Kato, M. Hirano, M. Onishi, and M. Nishimura, ''Ultra-low
nonlinearity low-loss pure silica core fibre for long-haul WDM
transmission,, Electron. Lett. {\bf 35} (19), 1615-1617 (1999).

\bibitem{hugo}  H. Zbinden, H. Bechmann-Pasquinucci, N. Gisin, G. Ribordy,
''Quantum Cryptography'', Appl. Phys. B {\bf 67}, 743-748 (1998).

\bibitem{B92}  Ch.H. Bennett, F. Bessette, G. Brassard, L. Salvail, and J.
Smolin, ''Experimental Quantum Cryptography'', J. Cryptology {\bf 5}, 3-28
(1992).

\bibitem{greg2000a}  G.Ribordy, J. Brendel, J.D. Gautier, N. Gisin, and H.
Zbinden, ''Long distance entanglement based quantum key distribution'',
quant-ph 0008039, Phys. Rev. A 63, 012309 (2001)

\bibitem{stucki2001}  D. Stucki, G. Ribordy, A. Stefanov, H. Zbinden, J.G.
Rarity, T. Wall, ''Photon counting for quantum key distribution with Peltier
cooled InGaAs APD's'', J. Mod. Opt, this issue (2001)

\bibitem{Brassard2}  G. Brassard, and L. Salvail, ''Secret-key
reconciliation by public discussion'', in {\it Advances in Cryptology,
Eurocrypt '93,} Lecture Notes in Computer Science\ {\bf 765}, 410-423 (1993).

\bibitem{}  Ch.H. Bennett, G. Brassard, C. Crepeau, and U.M. Maurer,
''Generalized privacy amplification'', IEEE Trans. Information th., {\bf 41}%
, 1915-1923 (1995).

\bibitem{greg2000b}  G. Ribordy, J.-D. Gautier, N. Gisin, O. Guinnard, H.
Zbinden, ''Fast and user-friendly quantum key distribution'', J. Mod. Opt., 
{\bf 47}, 517-531(2000).

\bibitem{Brassard}  G. Brassard, N. L\"{u}tkenhaus, T. Mor, and B.C.
Sanders, ''Limitations on Practical Quantum Cryptography'', Phys. Rev. Lett. 
{\bf 85}, 1330-1333 (2000).
\end{references}
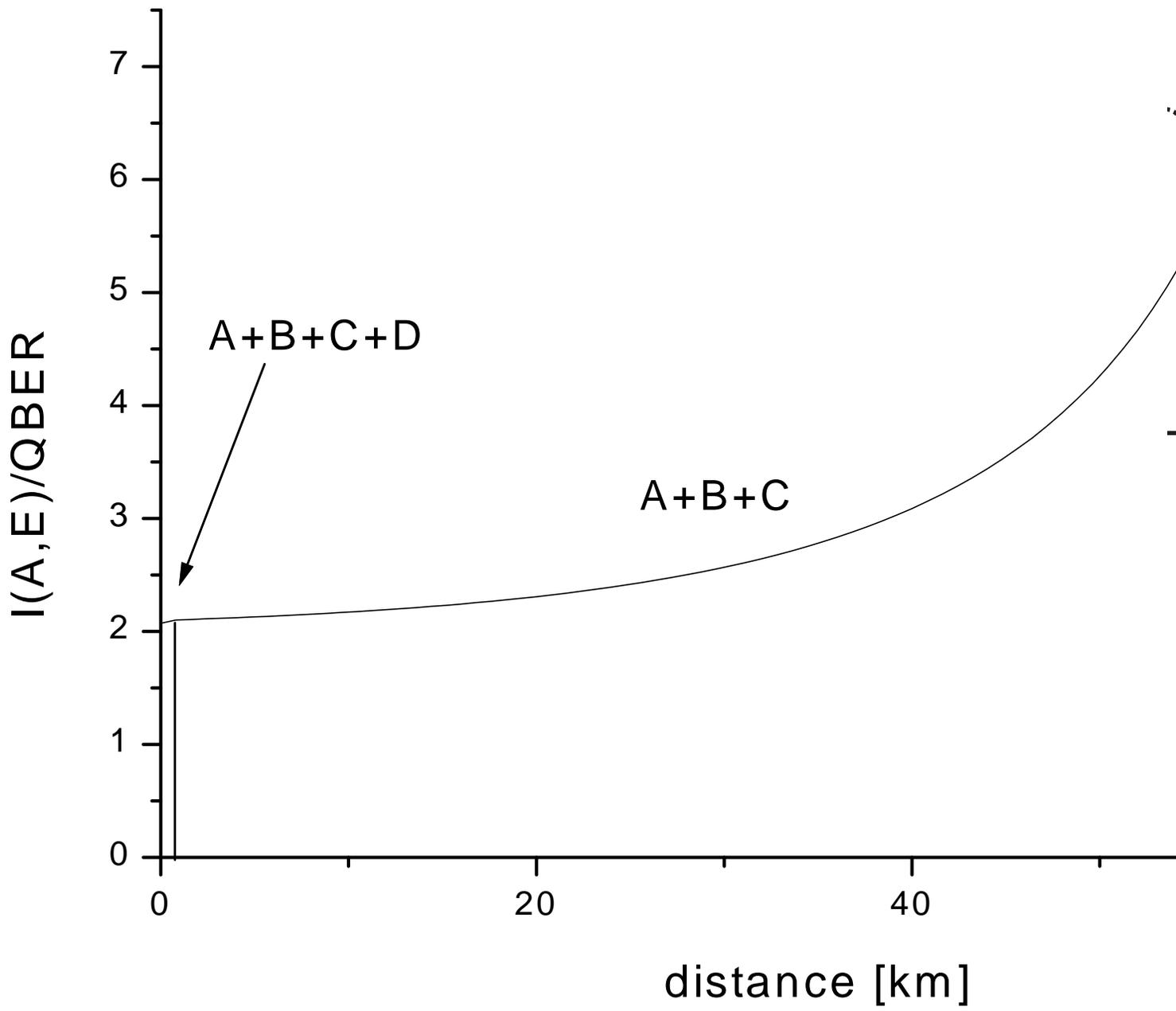
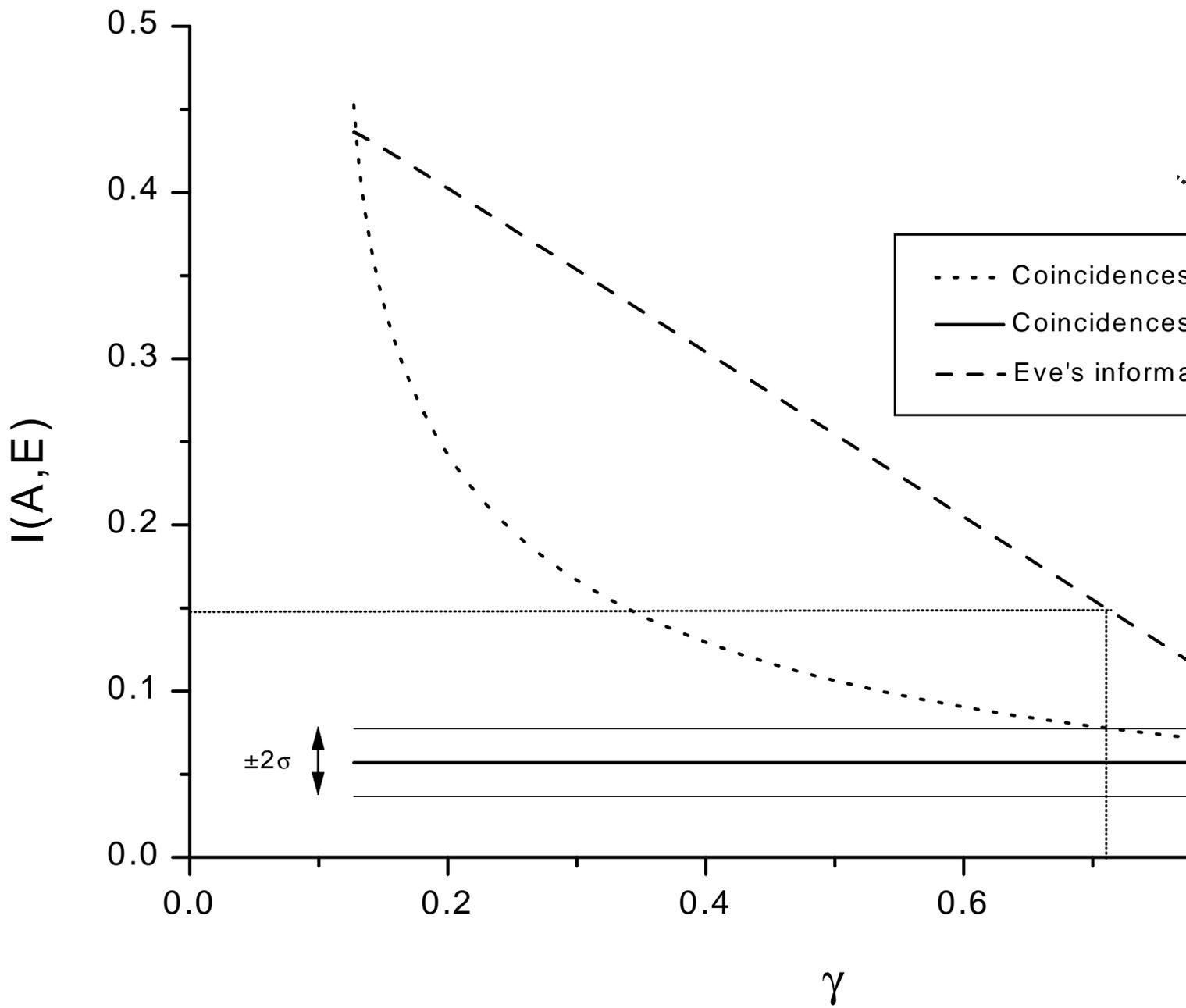
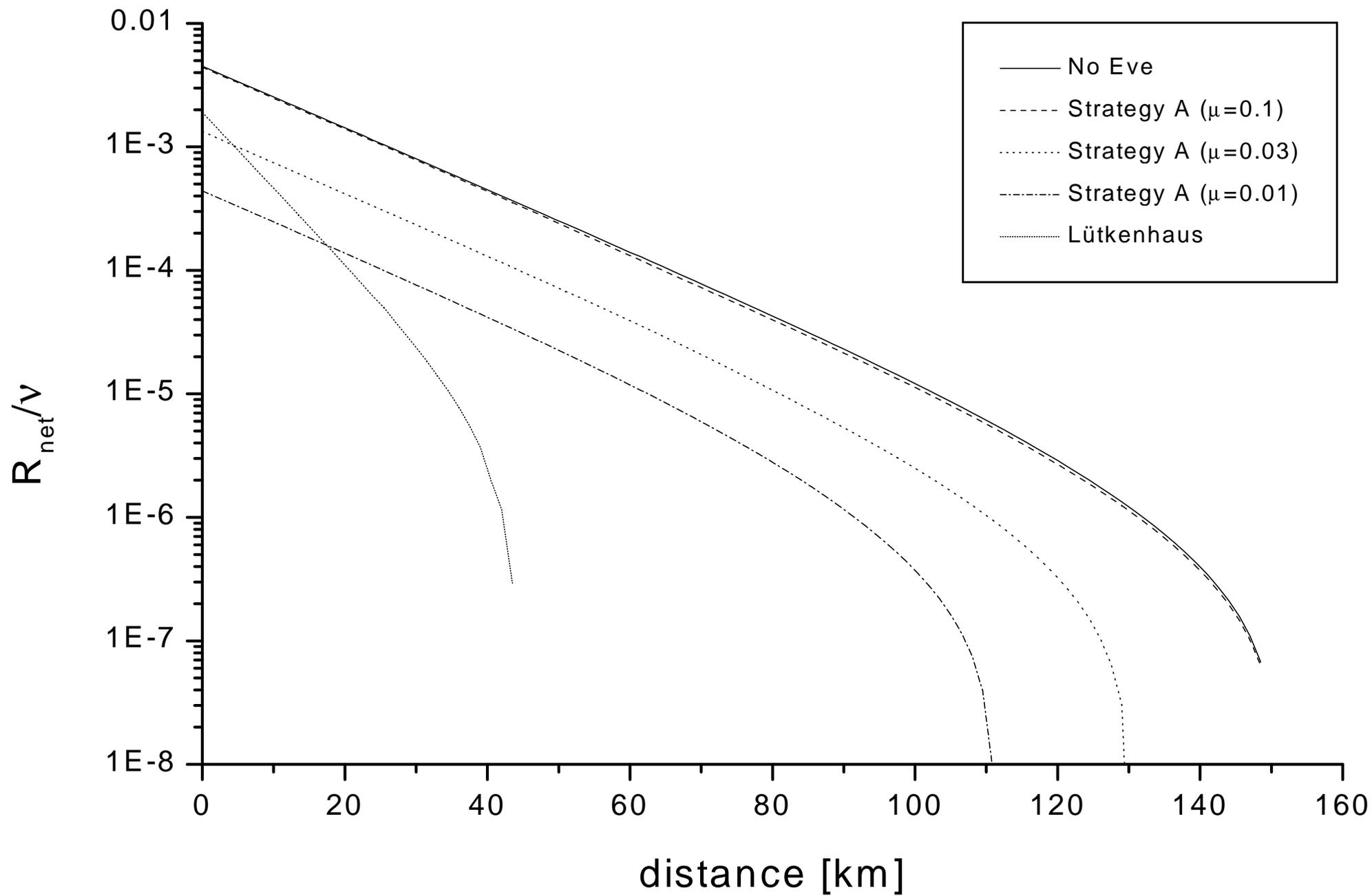
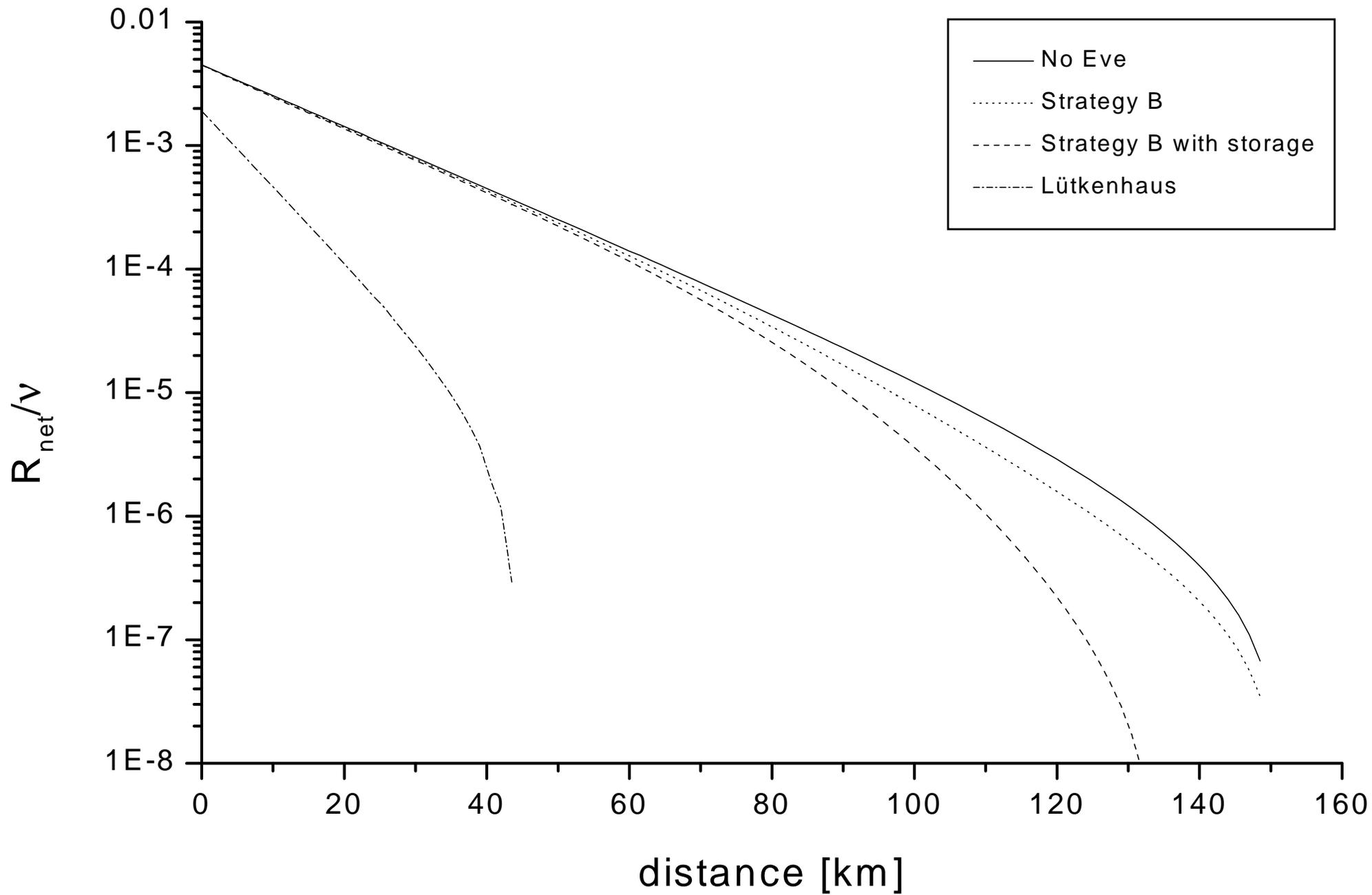
\end{document}